# Coevolution of Resource and Strategies in Common-Pool Resource Dilemmas: A Coupled Human-Environmental System Model

**Authors:** Chengyi Tu[1,2*], Renfei Chen[3], Ying Fan[4], Yongliang Yang[2]

[1]Department of Environmental Science, Policy, and Management, University of California, Berkeley; Berkeley, 94720, USA.

[2]School of Economics and Management, Zhejiang Sci-Tech University; Hangzhou, 310018, China.

[3]School of Life Science, Shanxi Normal University; Taiyuan, 030000, China.

[4]College of Geography and Environment, Shandong Normal University; Jinan, 250358, China.

*Corresponding author. Email: chengyitu1986@gmail.com

# Abstract

Common-pool resource governance requires users to cooperate and avoid overexploitation, but defection and free-riding often undermine cooperation. We model a human-environmental system that integrates dynamics of resource and users' strategies. The resource follows a logistic function that depends on natural growth rate, carrying capacity, and extraction rates of cooperators and defectors. The users' strategies evolve according to different processes that capture effects of payoff, resource, and noise. We analyze the feedback between resource availability and strategic adaptation, and explores the conditions for the emergence and maintenance of cooperation. We find different processes lead to different regimes of equilibrium solutions and resource levels depending on the parameter configuration and initial conditions. We also show that some processes can enhance the sustainability of the resource by making the users more responsive to the resource scarcity. The paper advances the understanding of human-environmental system and offers insights for resource governance policies and interventions.

# 1. Introduction

Common-pool resources (CPRs), such as fisheries, forests, and water, pose a significant challenge for human societies, as they are subject to overexploitation and depletion by multiple users (Davis et al., 2014; Dietz et al., 2003; Ostrom, 2008, 2009a; Suweis et al., 2015). CPRs are characterized by two features: non-excludability, which means that potential users cannot be easily prevented from accessing the resource, and subtractability, which means that each user's consumption reduces the resource availability for others (Ostrom, 1990; Van Laerhoven and Ostrom, 2007). These features create a social dilemma, where users face a trade-off between cooperation and defection (Ostrom, 1990; Tavoni et al., 2012). Cooperation can ensure the sustainability of the resource, but defection can yield higher individual payoffs, leading to a tragedy of the commons (Hardin, 1968; Hardin, 1998). To avoid this outcome, CPR users need to coordinate their actions and establish effective governance mechanisms (Ostrom, 2008).

The governance of CPRs has been greatly advanced by the seminal work of Elinor Ostrom, who proposed a framework of design principles for successful CPR institutions (Hauser et al., 2014; Ostrom, 1990, 2006, 2008, 2009a, b; Ostrom et al., 1994; Van Laerhoven and Ostrom, 2007). Ostrom showed that CPR users can achieve collective action through self-organization and self-governance, without relying on external authorities or privatization. She identified eight design principles that are common among long-enduring CPR institutions, such as clear boundaries, collective-choice arrangements, monitoring, graduated sanctions, conflict-resolution mechanisms, recognition of rights to organize, nested enterprises, and adaptation to local conditions. Ostrom's framework has been widely applied and tested in various empirical contexts, and has inspired numerous theoretical and experimental studies.

The study of CPR dilemmas has also been enriched by the analysis of various factors that can affect the level of cooperation among CPR users, such as social norms, institutions, communication, reputation, and punishment (Dietz et al., 2003; Hardin, 1968; Ostrom, 2009a, b; Van Laerhoven and Ostrom, 2007). However, most of these studies have neglected the endogenous and interdependent nature of the CPR dynamics, assuming that the resource availability is fixed and exogenous (Hilbe et al., 2018). This assumption overlooks that CPR users are not passive agents who react to a given environment, but active agents who adapt to and modify the environment. The resource availability influences not only the users' payoffs and incentives, but also their strategic choices and behaviors, and thus the dynamics and outcomes of the system (Liu et al., 2015; Turner et al., 2003).

A more realistic and comprehensive approach to CPR dilemmas is to consider the feedback between resource availability and strategic adaptation, and to model the CPR dynamics as endogenous and interdependent (Rocha et al., 2020; Runyan et al., 2015; Sethi and Somanathan, 1996). This approach elucidates the complex and non-linear interactions between users of CPR and the resource. It provides insights into how various factors, such as resource characteristics, user heterogeneity, learning processes, and institutional arrangements, can affect the emergence and evolution of cooperation (Schnellenbach and Schubert, 2014; Seabright, 1993; Van Laerhoven and Ostrom, 2007). Furthermore, this approach offers valuable perspectives on effective policies and interventions to enhance cooperation and sustainability in CPR dilemmas (Horan et al., 2011; Ostrom, 2009a; Schlüter and Pahl-Wostl, 2007; Turner et al., 2003; Wang and Fu, 2020).

In this paper, we present a coupled human-environmental system (HES) model that integrates the evolutionary dynamics of the CPR and the players' strategies. The CPR follows a logistic function that depends on the natural growth rate, the carrying capacity, and the extraction rates of cooperators and defectors. The players' strategies evolve according to different processes that capture the effects of payoff, resource, and noise. We analyze the feedback between CPR availability and strategic adaptation, and explore the conditions for the emergence and maintenance of cooperation in complex adaptive systems. We show that different processes lead to varying regimes of equilibrium solutions and CPR levels depending on the system parameter configuration and initial condition. We also discuss limitations and new avenues for future research on CPR governance using HES models and the implications of our results for CPR management and policy design.

## 2. Materials and methods

### 2.1 Evolutionary dynamics of resource

Building upon existing research, we construct a model that encapsulates the dynamic behavior of CPR, represented by $R(t)$, which follows a logistic function and is subject to extraction by $N$ players (Tilman, 2023; Tu et al., 2022). These players can choose between two strategies: cooperation, which involves sustainable resource extraction, or defection, which involves a larger and unsustainable level of extraction. The proportion of cooperators in the system is represented by $x = \frac{N_C}{N}$, where $N_C$ is the number of cooperators (and the proportion of defectors is $1-x$). The dynamics of the CPR follow a logistic function $TR(t)\left(1-\frac{R(t)}{K}\right)$, where $T > 0$ represents the natural growth rate and $K$ is the carrying capacity (Nelder, 1961; Tsoularis and Wallace, 2002; Verhulst, 1838). The extraction rates for cooperators and defectors are denoted by $e_C$ and $e_D$, respectively, with the condition that $0 < Ne_C < T < Ne_D$. The total extraction rate by all players is given by $E = N_C e_C + N_D e_D = N\left(xe_C + (1-x)e_D\right)$. Consequently, the differential equation for the resource volume of CPR can be expressed as follows:

$$\begin{aligned}
\frac{dR(t)}{dt} &= TR(t)\left(1-\frac{R(t)}{K}\right) - R(t)E = TR(t)\left(1-\frac{R(t)}{K}\right) - NR(t)\left(x(t)e_C + (1-x(t))e_D\right) \\
&= T\left(R(t)\left(1-\frac{R(t)}{K}\right) - R(t)\left(x(t)\hat{e}_C + (1-x(t))\hat{e}_D\right)\right)
\end{aligned} \quad (1)$$

where $\hat{e}_C = \frac{Ne_C}{T}, \hat{e}_D = \frac{Ne_D}{T}$ are normalized extraction parameters and $K=1$ by normalizing the resource volume $R$ between 0 and 1 for simplicity.

### 2.2 Evolutionary dynamics of players' strategies

In the scenario where players interact through a complete network, meaning each node is connected to all other nodes, the evolution of the probability $P^\tau(N_C)$, which represents the likelihood of having $N_C$ cooperators at time $\tau$, is dictated by following Master Equation (Traulsen et al., 2006; Traulsen and Hauert, 2009)

$$P^{\tau+1}(N_C) - P^{\tau}(N_C) = P^{\tau}(N_C-1)T^{D \to C}(N_C-1|R;\tau) + P^{\tau}(N_C+1)T^{C \to D}(N_C+1|R;\tau)$$
$$-P^{\tau}(N_C)T^{C \to D}(N_C|R;\tau) - P^{\tau}(N_C)T^{D \to C}(N_C|R;\tau)$$

where $T^{D \to C}(N_C \pm 1|R;\tau)$ is transition probability at time $\tau$ from $N_C$ to $N_C \pm 1$ given a resource level $R$.

Within a traditional evolutionary game framework, we can derive a generalized Fokker-Planck equation for the probability density $\rho(x)$ of observing the cooperators fraction $x$. This derivation begins with the Master Equation, and involves the substitutions $x = \frac{N_C}{N}, t = \frac{\tau}{N}$ and $\rho(x;t) = NP^{\tau}(N_C)$, leading to the following equation:

$$\rho(x;t+N^{-1}) - \rho(x;t) = \rho(x-N^{-1};t)T^{D \to C}(x-N^{-1}|R;t) + \rho(x+N^{-1};t)T^{C \to D}(x+N^{-1}|R;t)$$
$$-\rho(x;t)T^{C \to D}(x|R;t) - \rho(x;t)T^{D \to C}(x|R;t)$$

In the case where $N$ is significantly larger than 1, we can employ a Taylor series expansion around $x(t)$ to approximate both the probability densities and the transition probabilities. By disregarding terms of higher order in $N^{-1}$, we can derive the Fokker-Planck equation as follows:

$$\frac{d}{dt}\rho(x;t) = -\frac{d}{dx}[a(x|R;t)\rho(x;t)] + \frac{1}{2}\frac{d^2}{dx^2}[b^2(x|R;t)\rho(x;t)]$$

where $a(x|R;t) = T^+(x|R;t) - T^-(x|R;t)$ and $b(x|R;t) = \sqrt{\frac{1}{N}(T^+(x|R;t) + T^-(x|R;t))}$.

Given that the time steps in this model are independent, resulting in uncorrelated noise over time, we can utilize Ito calculus to derive the corresponding Langevin equation as follows:

$$\frac{dx(t)}{dt} = a(x|R;t) + b(x|R;t)\zeta$$

where $\zeta$ is an uncorrelated Gaussian white noise. In the limit as $N \to \infty$, the diffusion term $b(x|R;t)$ tends to zero with a rate of $1/\sqrt{N}$, resulting in a deterministic equation. Consequently, the evolutionary dynamics of the players' strategies, under the influence of the replicator rule, can be expressed as follows:

$$\frac{dx(t)}{dt} = T^+ - T^- \quad (2)$$

where $T^+ = T^{D \to C}, T^- = T^{C \to D}$ is transition probability for Master Equation.

By using different transition probabilities driven by replicator process, Moran process, and Fermi process, linear function of CPR, unit-step function of CPR, or logistic function of CPR, we can integrate the evolutionary dynamics of the CPR with the evolutionary dynamics of the players' strategies and are able to observe a variety of patterns (see Fig. 1). This comprehensive approach allows us to explore the complex interplay between resource availability and strategic decision-making in a competitive environment.

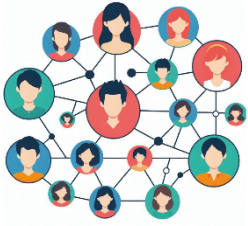 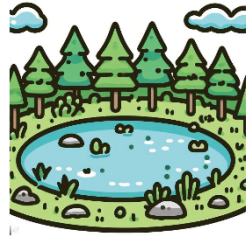 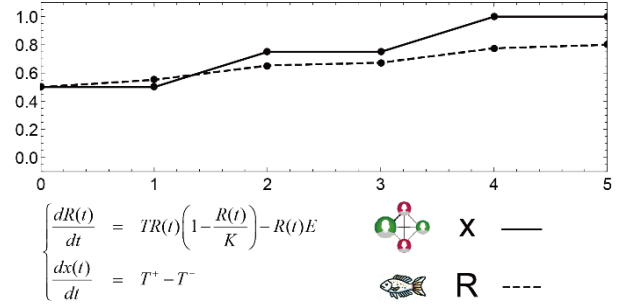

Evolutionary dynamics of players' strategies

$$\frac{dx(t)}{dt} = T^+ - T^-$$

Evolutionary dynamics of resource

$$\frac{dR(t)}{dt} = TR(t)\left(1 - \frac{R(t)}{K}\right) - R(t)E$$

Human-environmental system that integrates them

$$\begin{cases} \frac{dR(t)}{dt} = TR(t)\left(1 - \frac{R(t)}{K}\right) - R(t)E \\ \frac{dx(t)}{dt} = T^+ - T^- \end{cases}$$

**Fig. 1 | An illustration of human-environmental system (HES) model that integrates the evolutionary dynamics of the common-pool resource (CPR) and the players' strategies.**

## 2.3 Transition probability driven by replicator process

We start with the simplest case where the transition probability of the Master Equation determined by the replicator process, which is a model of how the frequencies of different individuals in a population evolve over time based on their interactions and payoffs (Schuster and Sigmund, 1983; Smith, 1982). The replicator process assumes that individuals are fixed and inherited, and that the fitness of each individual depends on its relative frequency in the population. The frequency of each individual changes proportionally to the difference between its fitness and the mean fitness of the population.

In our framework, the update rule of transition probability driven by the replicator process is as follows: 1) a random player $i$ is selected; 2) a random neighbor of $i$ is selected, $j$; 3) the player $i$ adopts the strategy of player $j$ with probability $p = \frac{1}{2} + \frac{w}{2}\frac{U_j - U_i}{\Delta U_{max}}$ where $U_i, U_j$ are the payoffs of players $i$ and $j$, respectively, $\Delta U_{max}$ is the maximum payoff difference that ensures $0 \leq p \leq 1$ ( $\Delta U_{max} = \max |U_D - U_C| = e_D - e_C$ when $R = 1$ ) and $-1 \leq w \leq 1$ is a greed parameter. The greed parameter $w$ measures the degree of greediness of a player, who prefers to defect and extract more resource than the cooperative and sustainable level. The cooperators fraction $x$ decreases as the greed parameter $w$ increases, leading to higher resource extraction and lower sustainability. Conversely, $x$ increases as the greed parameter $w$ decreases, leading to lower resource extraction and higher sustainability. In this paper, we focus on the sustainability of the resource, so we assume that $w$ is negative, i.e., $-1 \leq w < 0$. In this setting, the probabilities to switch strategy from defection to cooperation, and vice versa are $p^{D \rightarrow C} = \frac{1}{2} + \frac{w}{2}\frac{Re_C - Re_D}{e_D - e_C} = \frac{1}{2} - \frac{w}{2}R$ and $p^{C \rightarrow D} = \frac{1}{2} + \frac{w}{2}\frac{Re_D - Re_C}{e_D - e_C} = \frac{1}{2} + \frac{w}{2}R$, respectively, and the transition probabilities are $T^{D \rightarrow C} = x(1-x)p^{D \rightarrow C}$ and $T^{C \rightarrow D} = x(1-x)p^{C \rightarrow D}$. The evolutionary dynamics of players' strategies where the transition probability driven by

replicator process is

$$\frac{dx(t)}{dt} = T^{D \to C} - T^{C \to D} = -wR(t)(1-x(t))x(t) \quad (3)$$

## 2.4 Transition probability driven by Moran process

We then consider the Moran process, which is the global version of replicator process. The Moran process can be used to study evolutionary games, where the payoff of each individual is determined by the strategies of others in the population (Blume, 1993; Hauert and Szabó, 2005; Nowak et al., 2004; Szabó and Tőke, 1998; Taylor et al., 2004). In this process, the selection is modelled by assigning a relative fitness to each strategy, which reflects the probability of being chosen for reproduction. The same individual can be selected for both death and reproduction in the same time step, resulting in a stochastic birth-death process.

In our framework, the update rule of transition probability driven by the Moran process is as follows: 1) a random player $i$ is selected; 2) a random neighbor of $i$ is selected, $j$; 3) player $i$ adopts the strategy of player $j$ with probability $p = \frac{1-w+wU_j}{1-w+w\langle U \rangle}$ where $U_j$ are the payoff of player $j$, $\langle U \rangle = xU_C + (1-x)U_D = xRe_C + (1-x)Re_D$ and $-1 \leq w < 0$ is the greed parameter. In this setting, the probabilities to switch strategy from defection to cooperation, and vice versa are $p^{D \to C} = \frac{1-w+wU_C}{1-w+w\langle U \rangle}$ and $p^{C \to D} = \frac{1-w+wU_D}{1-w+w\langle U \rangle}$, respectively, and the transition probabilities are $T^{D \to C} = x(1-x)p^{D \to C}$ and $T^{C \to D} = x(1-x)p^{C \to D}$. The evolutionary dynamics of players' strategies where the transition probability driven by the Moran process is

$$\frac{dx(t)}{dt} = T^{D \to C} - T^{C \to D} = wRx(t)(1-x(t))\frac{e_C - e_D}{1-w+w(x(t)Re_C + (1-x(t))Re_D)} \quad (4)$$

## 2.5 Transition probability driven by Fermi process

We then explore the Fermi process, which is a hybrid model of population dynamics that combines the features of the replicator process and the Moran process. The Fermi process is a type of local update process, where individuals only interact with their neighbors on a network (Blume, 1993; Hauert and Szabó, 2005; Szabó and Tőke, 1998). However, unlike the replicator process, the Fermi process takes into account the external factors that affect the reproductive fitness of individuals, such as environmental noise or genetic variation. In this scenario, the game interaction payoff is only one of many additive components of the fitness function. When the external components are large and random, they act like a thermal noise that influences the strategy update. In this case, the probability of an individual adopting the strategy of its neighbor depends on the difference of their payoffs and a parameter that measures the intensity of the noise, analogous to the inverse temperature in statistical physics.

In our framework, the update rule of transition probability driven by the Fermi process is as follows: 1) a random player $i$ is selected; 2) a random neighbor of $i$ is selected, $j$; 3) player $i$ adopts the strategy of player $j$ with probability $p = \frac{1}{1+e^{-w(U_i - U_j)}}$ where $U_i, U_j$ are the payoffs of players $i$ and $j$, respectively, and $-1 \leq w < 0$ is the greed parameter. In this setting, the probabilities to switch strategy from defection to cooperation, and vice versa are $p^{D \to C} = \frac{1}{1+e^{-w(U_C - U_D)}}$ and $p^{C \to D} = \frac{1}{1+e^{-w(U_D - U_C)}}$, respectively, and the transition probabilities are $T^{D \to C} = x(1-x)p^{D \to C}$ and $T^{C \to D} = x(1-x)p^{C \to D}$. The evolutionary dynamics of players' strategies where the transition probability driven by Fermi process is

$$\frac{dx(t)}{dt} = T^{D \to C} - T^{C \to D} = x(t)(1-x(t))\tanh\left(\frac{w}{2}(Re_C - Re_D)\right) \quad (5)$$

## 2.6 Transition probability driven by linear function of CPR

We consider a realistic scenario where the players' decisions are influenced by the CPR availability, which reflects the environmental conditions and the collective actions of the players. The CPR availability affects both the players' strategies and payoffs, and thus the dynamics and outcomes of the system. To capture this effect, we introduce a transition probability driven by a linear function of CPR, which determines the likelihood of a player switching from cooperation to defection or vice versa based on the current level of CPR in the system. This implies that a player is more inclined to cooperate when the CPR is scarce and defect when the CPR is abundant, in order to sustain the CPR. We investigate how this parameter affects the stability and sustainability of the system, and show that it can result in different regimes of equilibrium solutions and CPR levels depending on the parameter values and initial conditions of the system.

In our framework, the update rule of transition probability driven by the linear function of CPR is as follows: 1) a random player $i$ is selected; 2) player $i$ updates its strategy according to the resource availability. If the player $i$ is cooperator, it switches to defector with probability $p^{C \to D} = p = R$, and if the player $i$ is defector, it switches to cooperator with probability $p^{D \to C} = 1 - p$. In this setting, the probabilities to switch strategy from cooperation to defection, and vice versa are $p^{C \to D} = p = R$ or $p^{D \to C} = 1 - p$, respectively, and the transition probabilities are $T^{D \to C} = (1-x)p^{D \to C}$ and $T^{C \to D} = xp^{C \to D}$. The evolutionary dynamics of players' strategies where transition probability driven by linear function with resource is

$$\frac{dx(t)}{dt} = T^{D \to C} - T^{C \to D} = 1 - x - p = 1 - x - R \quad (6)$$

## 2.7 Transition probability driven by unit-step function of CPR

We investigate a scenario in which players behave rationally and switch between cooperation and defection according to a threshold value of the CPR. The threshold value reflects the minimum level of cooperation needed to sustain the CPR. The switching probability is determined by a unit-step function of the threshold value. Our focus is how the transition probabilities between strategies are affected by the interplay of the unit-step function and CPR. This framework provides a novel and simple perspective on the feedback between CPR availability and strategic adaptation, and enhances our understanding of complex adaptive systems.

In our framework, the update rule of transition probability driven by the unit-step function of CPR is as follows: 1) a random player $i$ is selected; 2) the player $i$ updates its strategy according to the and an offset parameter for resource availability. If the player $i$ is cooperator, it switches to defector with probability $p^{C \to D} = p = \theta[R-c]$ where $0 < c < 1$ is the threshold, and if the player $i$ is defector, it switches to cooperator with probability $p^{D \to C} = 1 - p$. In this setting, the probabilities to switch strategy from cooperation to defection, and vice versa are $p^{C \to D} = p = \theta[R-c]$ and $p^{D \to C} = 1 - p$, respectively, and the transition probabilities are $T^{D \to C} = (1-x)p^{D \to C}$ and $T^{C \to D} = xp^{C \to D}$. The evolutionary dynamics of players' strategies where the transition probability driven by unit-step function of CPR is

$$\frac{dx(t)}{dt} = T^{D \to C} - T^{C \to D} = 1 - x - p = 1 - x - \theta[R-c] \quad (7)$$

## 2.8 Transition probability driven by logistic function of CPR

We study a scenario in which players behave rationally and adjust their strategies of cooperation and defection based on a logistic function of the CPR (Berkson, 1944; Jordan, 1995). The logistic function captures the trade-off between the cooperation and defection as well as the non-linear effects of CPR depletion/regeneration. Our aim is how the transition probabilities between strategies are influenced by the shape and parameters of the logistic function and CPR dynamics. This framework offers a novel and general perspective on the feedback between CPR availability and strategic adaptation, and improves our understanding of complex adaptive systems.

In our framework, the update rule of transition probability driven by the logistic function with resource is as follows: 1) a random player $i$ is selected; 2) the player $i$ updates its strategy according to the and an offset parameter for resource availability. If the player $i$ is cooperator, it switches to defector with probability $p^{C \to D} = p = \dfrac{1}{1 + \exp^{-k(R-c)}}$ where $k > 0$ is the intensity parameter and $0 < c < 1$ is the threshold parameter, and if the player $i$ is defector, it switches to cooperator with probability $p^{D \to C} = 1 - p$. In this setting, the probabilities to switch strategy from cooperation to defection, and vice versa are $p^{C \to D} = p = \dfrac{1}{1+\exp^{-k(R-c)}}$ and $p^{D \to C} = 1-p$, respectively, and the transition probabilities are $T^{D \to C} = (1-x)p^{D \to C}$ and $T^{C \to D} = xp^{C \to D}$. The evolutionary dynamics of players' strategies where the transition probability driven by logistic function of CPR is

$$\frac{dx(t)}{dt} = T^{D \to C} - T^{C \to D} = 1 - x - p = 1 - x - \frac{1}{1+\exp^{-k(R-c)}} \quad (8)$$

In conclusion, we present a comprehensive overview of the six distinct transition probabilities and their associated evolutionary dynamics in relation to players' strategies, as detailed in Tab. 1.

**Tab. 1 | An overview of the six transition probabilities and their evolutionary dynamics of players' strategies.**

| Driver | Evolutionary dynamics of players' strategies |
|---|---|
| replicator process | $\dfrac{dx(t)}{dt} = -wR(t)(1-x(t))x(t)$ |
| Moran process | $\dfrac{dx(t)}{dt} = wRx(t)(1-x(t)) \dfrac{e_C - e_D}{1-w+w(x(t)Re_C + (1-x(t))Re_D)}$ |
| Fermi process | $\dfrac{dx(t)}{dt} = x(t)(1-x(t))\tanh\left(\dfrac{w}{2}(Re_C - Re_D)\right)$ |
| linear function of CPR | $\dfrac{dx(t)}{dt} = 1 - x - R$ |
| unit-step function of CPR | $\dfrac{dx(t)}{dt} = 1 - x - \theta[R-c]$ |
| logistic function of CPR | $\dfrac{dx(t)}{dt} = 1 - x - \dfrac{1}{1+\exp^{-k(R-c)}}$ |

# 3. Results

We construct a coupled human-environmental system (HES) model that integrates the evolutionary dynamics of the CPR and the players' strategies. The transition probability between strategies is driven by various processes that capture the effects of payoff, resource, and noise. We obtain the full dynamics of the HES model by combining the equations for the CPR and the strategies. This framework allows us to analyse the feedback between CPR and strategic adaptation, and to explore the conditions for the emergence and maintenance of cooperation in complex adaptive systems.

## 3.1 Replicator process-driven transition probability in coevolutionary system

By integrating the evolutionary dynamics of CPR, Eq. (1), with the evolutionary dynamics of players' strategies where the transition probability is driven by the replicator process, Eq. (3), we derive the following set of coupled equations:

$$\begin{cases} \dfrac{dR(t)}{dt} = T\left(R(t)\left(1-\dfrac{R(t)}{K}\right) - R(t)\left(x(t)\hat{e}_C + (1-x(t))\hat{e}_D\right)\right) \\ \dfrac{dx(t)}{dt} = -wR(t)x(t)(1-x(t)) \end{cases} \quad (9)$$

We analyze the stationary solutions of the coupled equations $\{R^*, x^*\}$, which are given by $s_1 = \{R=0, \forall x\}$ where the system depletes the CPR and $s_2 = \{R=1-\hat{e}_C, x=1\}$ where the system sustains the CPR. We then examine the stability of these solutions for the parameter configuration $T, \hat{e}_C, \hat{e}_D, w$ and initial condition $R_0, x_0$. The stability of these solutions cannot be easily assessed, but we can estimate it by the determinant and trace of the two-dimensional Jacobian matrix. If the determinant is positive and the trace is negative, i.e., $\text{Det} > 0 \wedge \text{Tr} < 0$, the solution is stable. Using this method, we determine that $s_1$ is neutral stable for $x < \dfrac{-1+\hat{e}_D}{-\hat{e}_C+\hat{e}_D}$ and $s_2$ is stable. Therefore, the system exhibits bi-stability depending on the parameter configuration $T, \hat{e}_C, \hat{e}_D, w$ and initial condition $R_0, x_0$: one stable state $s_2$ with CPR sustainability and full cooperator fraction and another stable state $s_1$ with CPR depletion and low cooperator fraction (see Fig. 2 a).

## 3.2 Moran process-driven transition probability in coevolutionary system

By integrating the evolutionary dynamics of CPR, Eq. (1), with the evolutionary dynamics of players' strategies where the transition probability is driven by the Moran process, Eq. (4), we derive the following set of coupled equations:

$$\begin{cases} \dfrac{dR(t)}{dt} = T\left(R(t)(1-R(t)) - R(t)\left(x(t)\hat{e}_C + (1-x(t))\hat{e}_D\right)\right) \\ \dfrac{dx(t)}{dt} = wR(t)x(t)(1-x(t))\dfrac{e_C - e_D}{1-w+w(x(t)R(t)e_C + (1-x(t))Re_D)} \end{cases} \quad (10)$$

We analyze the stationary solutions of the coupled equations $\{R^*, x^*\}$, which are given by $s_1 = \{R=0, \forall x\}$ where the system depletes the CPR and $s_2 = \{R=1-\hat{e}_C, x=1\}$ where the system sustains the CPR. We then examine the stability of these solutions for the parameter configuration $T, \hat{e}_C, \hat{e}_D, w$ and initial condition $R_0, x_0$. Using the method of $\text{Det} > 0 \wedge \text{Tr} < 0$, we determine that $s_1$ is neutral stable for $x < \dfrac{-1+\hat{e}_D}{-\hat{e}_C+\hat{e}_D}$ and $s_2$ is stable. Therefore, the system exhibits similar bi-stability as the previous case of the replicator process (see Fig. 2 b).

## 3.3 Fermi process-driven transition probability in coevolutionary system

By integrating the evolutionary dynamics of CPR, Eq. (1), with the evolutionary dynamics of players' strategies where the transition probability is driven by the Fermi process, Eq. (5), we derive the following set of coupled equations:

$$\begin{cases} \dfrac{dR(t)}{dt} = T\Big(R(t)(1-R(t)) - R(t)\big(x(t)\hat{e}_C + (1-x(t))\hat{e}_D\big)\Big) \\ \dfrac{dx(t)}{dt} = x(t)(1-x(t))\tanh\left(\dfrac{w}{2}\big(R(t)e_C - R(t)e_D\big)\right) \end{cases} \quad (11)$$

We analyze the stationary solutions of the coupled equations $\{R^*, x^*\}$, given by $\mathbf{s}_1 = \{R=0, \forall x\}$ where the system depletes the CPR and $\mathbf{s}_2 = \{R=1-\hat{e}_C, x=1\}$ where the system sustains the CPR. We then examine the stability of these solutions for the parameter configuration $\hat{e}_C, \hat{e}_D$ and initial condition $R_0, x_0$. Using the method of $\text{Det} > 0 \wedge \text{Tr} < 0$, we determine that $\mathbf{s}_1$ is neutral stable for $x < \dfrac{-1+\hat{e}_D}{-\hat{e}_C + \hat{e}_D}$ and $\mathbf{s}_2$ is stable. Therefore, the system exhibits similar bi-stability as the previous cases of the replicator process and Moran process (see Fig. 2 c).

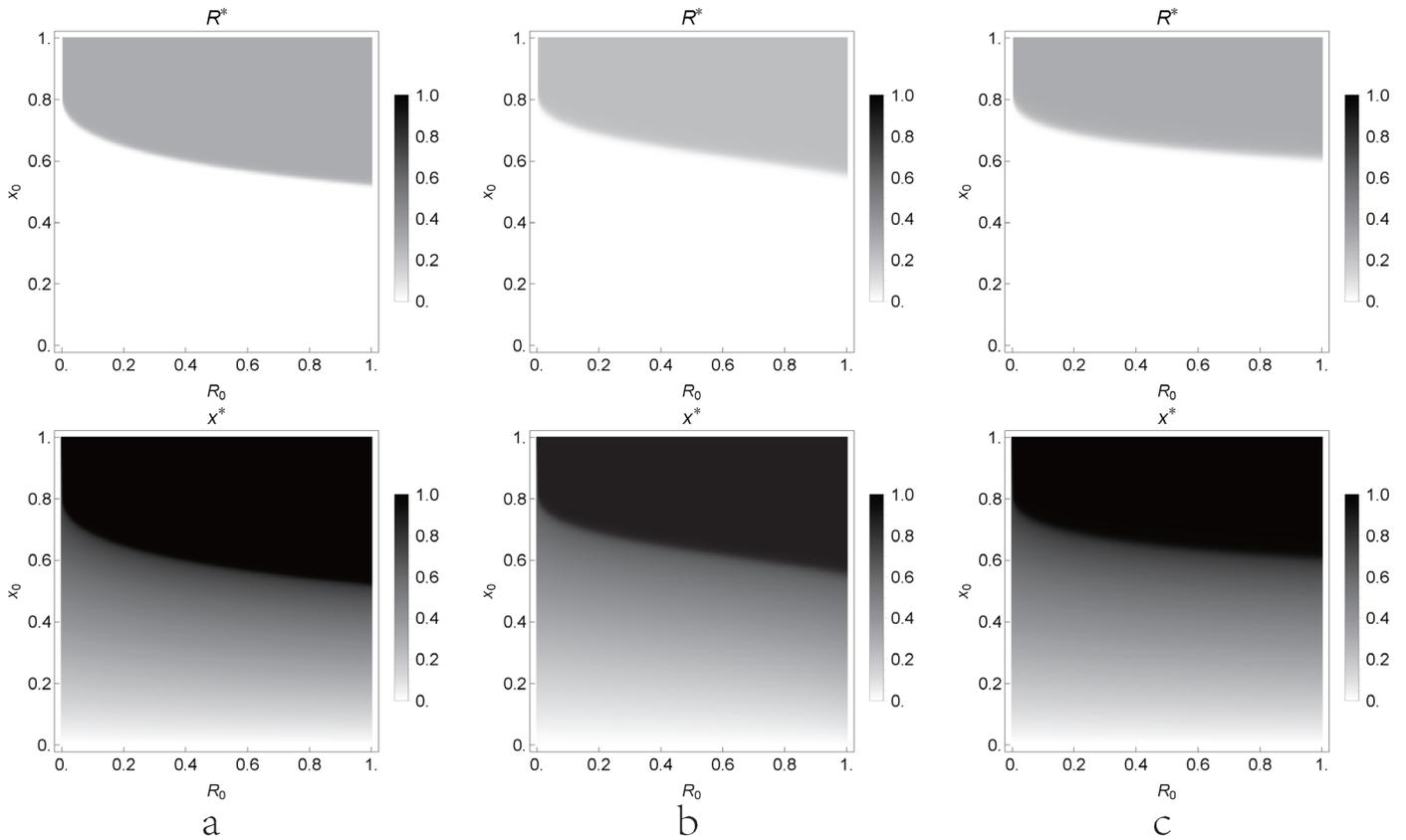

**Fig. 2 | The stable equilibrium of the coevolutionary system with different transition probabilities driven by the replicator process, Moran process, and Fermi process.** (a) Density plot of stable equilibrium $R^*$ and $x^*$ for replicator process. (b) Density plot of stable equilibrium $R^*$ and $x^*$ for Moran process. (c) Density plot of stable equilibrium $R^*$ and $x^*$ for

Fermi process. In each panel, greed parameter $w = -1$ and parameter configuration $T = 2, \hat{e}_C = 0.7, \hat{e}_D = 2$.

## 3.4 Linear function of CPR-driven transition probability in coevolutionary system

By integrating the evolutionary dynamics of CPR, Eq. (1), with the evolutionary dynamics of players' strategies where the transition probability is driven by the linear function of CPR, Eq. (6), we derive the following set of coupled equations:

$$\begin{cases} \dfrac{dR(t)}{dt} = T\Big(R(t)(1-R(t)) - R(t)\big(x(t)\hat{e}_C + (1-x(t))\hat{e}_D\big)\Big) \\ \dfrac{dx(t)}{dt} = 1 - x - R \end{cases} \quad (12)$$

We analyze the stationary solutions of the coupled equations $\{R^*, x^*\}$, given by $\mathbf{s}_1 = \{R = 0, x = 1\}$ where the system depletes the CPR and $\mathbf{s}_2 = \{R = \dfrac{1-\hat{e}_C}{1-\hat{e}_C + \hat{e}_D}, x = \dfrac{\hat{e}_D}{1-\hat{e}_C + \hat{e}_D}\}$ where the system sustains the CPR. We then examine the stability of these solutions for the parameter configuration $\hat{e}_C, \hat{e}_D$ and initial condition $R_0, x_0$. Using the method of $\mathrm{Det} > 0 \wedge \mathrm{Tr} < 0$, we determine that $\mathbf{s}_1$ is unstable and $\mathbf{s}_2$ is stable. Therefore, the CPR is maintained by the system regardless of the parameter configuration $\hat{e}_C, \hat{e}_D$ and initial condition $R_0, x_0$ (see Fig. 3 a).

## 3.5 Unit-step function of CPR-driven transition probability in coevolutionary system

By integrating the evolutionary dynamics of CPR, Eq. (1), with the evolutionary dynamics of players' strategies where the transition probability is driven by the unit-step function of CPR, Eq. (7), we derive the following set of coupled equations:

$$\begin{cases} \dfrac{dR(t)}{dt} = T\Big(R(t)(1-R(t)) - R(t)\big(x(t)\hat{e}_C + (1-x(t))\hat{e}_D\big)\Big) \\ \dfrac{dx(t)}{dt} = 1 - x - \theta[R - c] \end{cases} \quad (13)$$

We assume a symmetric unit-step function for the transition probability, where the threshold value $c = 0.5$, which corresponds to the midpoint of the resource range. We analyze the stationary solutions of the coupled equations $\{R^*, x^*\}$, given by $\mathbf{s}_1 = \{R = 0, x = 1\}$ where the system depletes the CPR and $\mathbf{s}_2 = \{R = 1 - \hat{e}_C, x = 1\}$ where the system sustains the CPR. We then examine the stability of these solutions for the parameter configuration $\hat{e}_C, \hat{e}_D$ and initial condition $R_0, x_0$. Using the method of $\mathrm{Det} > 0 \wedge \mathrm{Tr} < 0$, we determine that $\mathbf{s}_1$ is unstable and $\mathbf{s}_2$ is stable. As in the previous case, the CPR is maintained by the system regardless of the parameter configuration $\hat{e}_C, \hat{e}_D$ and initial condition $R_0, x_0$, although the equilibrium value is different (see Fig. 3 b).

## 3.6 Logistic function of CPR-driven transition probability in coevolutionary system

By integrating the evolutionary dynamics of CPR, Eq. (1), with the evolutionary dynamics of players' strategies where the transition probability is driven by the logistic function of CPR, Eq. (8), we derive the following set of coupled equations:

$$\begin{cases} \dfrac{dR(t)}{dt} = T\Big(R(t)\big(1-R(t)\big) - R(t)\big(x(t)\hat{e}_C + \big(1-x(t)\big)\hat{e}_D\big)\Big) \\ \dfrac{dx(t)}{dt} = 1 - x - \dfrac{1}{1+\exp^{-k(R-c)}} \end{cases} \quad (14)$$

We assume a symmetric and S-shaped logistic function for the transition probability, where the threshold $c = 0.5$ corresponds to the midpoint of the resource range and intensity $k$ controls the S-shaped. The transition probability driven by the logistic function depends on the intensity parameter $k$, which controls the steepness of the curve. As the intensity parameter $k$ increases, the logistic function approaches a unit-step function, which has a sudden jump from 0 to 1 at a certain point. As the intensity parameter $k$ decreases, the logistic function approaches a linear function, which has a constant slope between 0 and 1. Thus, the logistic function can be seen as a compromise between the unit-step function and the linear function. We encounter difficulties in analyzing the properties and stability of the solution analytically, but we use numerical simulations to explore the system's dynamics. We observe that, depending on the value of $k$, the system can either deplete or sustain the CPR, regardless of the parameter configuration $\hat{e}_C, \hat{e}_D$ and initial condition $R_0, x_0$ (see Fig. 3 c).

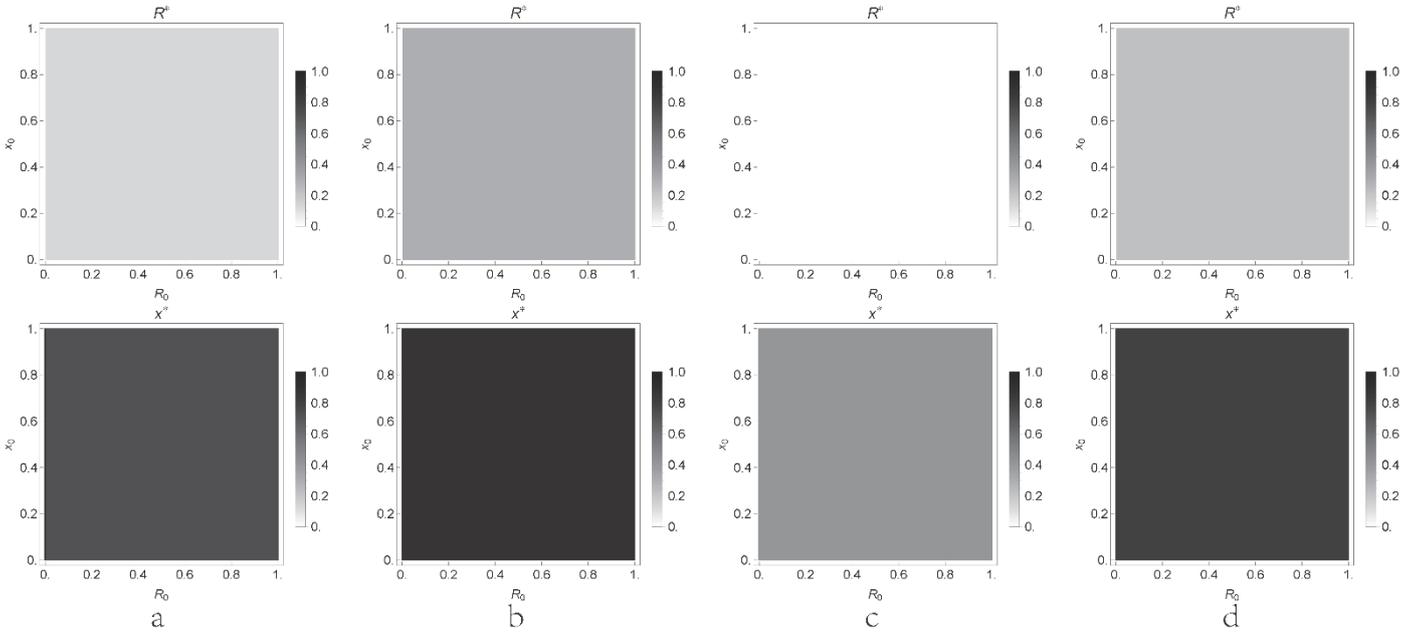

**Fig. 3 | The stable equilibrium of the coevolutionary system with different transition probabilities driven by the linear function with resource, unit-step function with resource, and logistic function of CPR.** (**a**) Density plot of stable equilibrium $R^*$ and $x^*$ for the linear function of CPR. (**b**) Density plot of stable equilibrium $R^*$ and $x^*$ for the unit-step function of CPR where $c = 0.5$. (**c**) Density plot of stable equilibrium $R^*$ and $x^*$ for the logistic function of CPR where $c = 0.5$ and $k = 0.1$. (**d**) Density plot of stable equilibrium $R^*$ and $x^*$ for the logistic function of CPR where $c = 0.5$ and $k = 10$. In each panel, parameter configuration $T = 2, \hat{e}_C = 0.7, \hat{e}_D = 2$.

Finally, we compare the effects of six distinct transition probabilities driven by the replicator process, Moran process, Fermi process, linear function of CPR, unit-step function of CPR, and logistic function of CPR (see Tab. 2). We find that the CPR system, where the transition probabilities driven by the replicator process, Moran process and Fermi process, exhibits two distinct regimes: a sustainable regime with high cooperation and resource levels, and a depleted regime with low cooperation and resource levels. Therefore, the transition probabilities based on evolutionary processes can lead to bi-stability, where the system can converge to either regime depending on the initial conditions and parameter values (see Fig. 2).

The transition probabilities based on resource-dependent functions can promote sustainability, where the system always converges to the sustainable regime regardless of the initial conditions and parameter values, provided that the intensity of the resource effect is sufficiently high, except when the intensity parameter $k$ is low for the logistic function of CPR (see Fig. 3). This indicates that the intensity parameter $k$ plays a crucial role in determining the outcome of the system. A low value of $k$ implies that the players are less sensitive to the resource level and more likely to defect, leading to a tragedy of the commons scenario. A high value of $k$ implies that the players are more sensitive to the resource level and more likely to cooperate, leading to a sustainable management scenario. Our results demonstrate that the transition probabilities between strategies are crucial for determining the fate of the CPR system, and that different mechanisms can result in different equilibrium outcomes and resource levels. Our results also highlight that resource availability can affect the players' choices and payoffs, and incorporating this feedback into the transition probabilities can enhance cooperation and sustainability.

**Tab. 2 | The stable equilibrium of the coevolutionary system with different transition probabilities.**

| Transition probability | Stable state | |
| --- | --- | --- |
| | Depleted state | Sustainable state |
| replicator process | $\mathbf{s}_1 = \{R = 0, \forall x\}$ | $\mathbf{s}_2 = \{R = 1 - \hat{e}_C, x = 1\}$ |
| Moran process | $\mathbf{s}_1 = \{R = 0, \forall x\}$ | $\mathbf{s}_2 = \{R = 1 - \hat{e}_C, x = 1\}$ |
| Fermi process | $\mathbf{s}_1 = \{R = 0, \forall x\}$ | $\mathbf{s}_2 = \{R = 1 - \hat{e}_C, x = 1\}$ |
| linear function of CPR | None | $\mathbf{s}_2 = \{R = \dfrac{1 - \hat{e}_C}{1 - \hat{e}_C + \hat{e}_D}, x = \dfrac{\hat{e}_D}{1 - \hat{e}_C + \hat{e}_D}\}$ |
| unit-step function of CPR | None | $\mathbf{s}_2 = \{R = 1 - \hat{e}_C, x = 1\}$ |
| logistic function of CPR | intensity parameter $k$ is small | intensity parameter $k$ is large |

# 4 Discussion and Conclusion

We present a novel framework for studying the coevolution of common-pool resources (CPR) and human strategies in complex adaptive systems. We use a coupled human-environmental system (HES) model that integrates the evolutionary dynamics of the resource and the players' strategies. The transition probability between strategies is driven by various processes that capture the effects of payoff, resource, and noise. We explore the conditions for the emergence and maintenance of cooperation and sustainability in CPR governance.

The main contribution of the paper is to provide a general and flexible approach to model the feedback between resource availability and strategic adaptation in CPR dilemmas. We show that the transition probability between strategies can be influenced by different factors, such as payoff, resource, noise, threshold, intensity, and offset. These factors can reflect the heterogeneity, rationality, and adaptability of the players, as well as the environmental conditions and collective actions of the system. We demonstrate that different transition probabilities can lead to varying outcomes of cooperation and sustainability in CPR governance. For example, we show that linear, unit-step, and logistic functions of CPR can result in CPR maintenance regardless of the parameter configuration and initial condition. In contrast, replicator, Moran, and Fermi processes can result in bi-stability depending on the parameter configuration and initial condition.

The paper has some limitations that could be addressed in future research. First, the paper assumes a homogeneous population of players with identical extraction rates and payoff functions. It would be interesting to extend the model to include heterogeneous players with different preferences, beliefs, and behaviors. Second, the paper focuses on a single CPR with a logistic growth function. It would be useful to generalize the model to multiple CPRs with different dynamics and interactions. Third, the paper uses a mean-field approximation to derive the evolutionary equations for the players' strategies. It would be important to validate the results with agent-based simulations or empirical data. Fourth, the paper does not consider the effects of social norms, institutions, or policies on CPR governance. It would be relevant to incorporate these factors into the model and analyze how they affect the coevolution of resource and strategies.

The paper opens up new avenues for further research on CPR governance using HES models. Some possible extensions include: exploring other types of transition probabilities that can capture more realistic and sophisticated decision-making processes; incorporating additional factors that can affect CPR dynamics, such as institutional arrangements, social norms, learning, and innovation; applying HES models to empirical case studies of CPR dilemmas, such as fisheries, forests, water, and climate; and developing methods and tools for analyzing and managing HES models, such as bifurcation analysis, resilience assessment, and adaptive management.

# Code and data availability

Code and data for all analyses are available at OSF (https://osf.io/b2vz6/).

# Declaration of Competing Interests

The author declares that there are no known competing financial interests or personal relationships that influenced the work reported in this paper.

# Acknowledgments

This work was supported by Microsoft AI for Earth, Experimental Social Science Laboratory at University of California Berkeley, Natural Science Fund of Zhejiang Province under Grant No. LZ18G010001, LZ22G010001, Science Foundation of Zhejiang Sci-Tech University under Grant No. 18092125-Y, 22092034-Y.